\documentclass[mathleft]{an}
\usepackage{graphicx,natbib,xspace}
\usepackage{times}
\let\jnl=\rmfamily
\def\refe@jnl#1{{\jnl#1}}%

\newcommand\aap{\refe@jnl{A\&A}}%
\newcommand\aapr{\refe@jnl{A\&A~Rev.}}%
\newcommand\aaps{\refe@jnl{A\&AS}}%
\newcommand\memras{\refe@jnl{MmRAS}}%
\newcommand\nat{\refe@jnl{Nature}}%

\newcommand{\about}{$\simeq$}

\newcommand{\Co}{$^{56}$Co\xspace} 

\newcommand{\Ni}{$^{56}$Ni\xspace}

\newcommand{\Msol}{M\ensuremath{_\odot}\xspace}

\begin{document}

\Pagespan{001}{}
\Yearpublication{2015}%
\Yearsubmission{2015}%
\Month{06}%
\Volume{336}%
\Issue{05}%
 \DOI{10.1002/asna.201512179}%

\title{Gamma rays from a supernova of type Ia: SN2014J}

\author{Roland Diehl\thanks{
  \email{rod@mpe.mpg.de}\newline}
  }
\titlerunning{Gamma rays from SN2014J}
\authorrunning{R. Diehl}
\institute{
Max-Planck-Institut f\"ur extraterrestrische Physik,
  D-85741 Garching, Germany
  \label{inst:mpe}
}

\received{15 Feb 2015}
\accepted{08 Apr 2015}
\publonline{22 Jun 2015}

\keywords{supernovae: individual -- supernovae: general -- gamma rays: observations -- techniques: spectroscopic -- nuclear reactions, nucleosynthesis, abundances}

\abstract{%
  SN2014J is the closest supernova of type Ia that occured in the last 40 years. This provides an opportunity for unprecedented observational detail and coverage in many astronomical bands, which will help to better understand the still unknown astrophysics of these supernovae. For the first time, such an event occurs sufficiently nearby so that also gamma rays are able to contribute to such investigations. This is important, as the primary source of  the supernova light is the radioactive energy from about 0.5 M$_\odot$ 
of $^{56}$Ni produced in the explosion, and the gamma rays associated with this decay make the supernova shine for months. The INTEGRAL gamma-ray observatory of ESA has followed the supernova emission for almost 5 months. The characteristic gamma ray lines from the $^{56}$Ni  decay chain through $^{56}$Co to  $^{56}$Fe have been measured. We discuss these observations, and the implications of the measured gamma-ray line characteristics as they evolve.}

\maketitle

\section{Introduction}
Supernovae of type Ia are commonly agreed to originate from the thermonuclear explosion of a CO white dwarf star in a binary system \citep{Hoyle:1960}. Accretion of material from the companion star onto the white dwarf eventually leads to nuclear ignition, and the nuclear energy release from carbon fusion ignited in the central degenerate region of the white dwarf occurs so fast that the white dwarf cannot adjust its structure by expansion, and rather is disrupted \citep[see, e.g.,][for a recent review]{Hillebrandt:2013aa}. Different paths have been proposed of how the supernova could be initiated from binary interaction  \citep[e.g.][]{Piersanti:2014}: (1) The mass of the white dwarf may be increased from accretion of material from the companion star to and above the Chandrasekhar mass limit of stability for degenerate matter; (2) a major disturbing event may occur on the white dwarf surface and cause the white dwarf interior to become unstable towards runaway nuclear carbon fusion, e.g. accretion of a major amount of mass in form of a colliding body or a material cloud, or a nuclear explosion of accreted helium on the surface. 

Nuclear fusion at high densities processes the white dwarf material to iron group nuclei, which are the most stable configuration of nuclear matter. \Ni is a likely product of such explosive supernova nucleosynthesis under typical central densities of white dwarfs of 10$^7$~g~cm$^{-3}$ or above and temperatures of a few GK \citep{Nomoto:1997aa}. The nuclear flame in principle propagates  through heat conduction, but is strongly accelerated through turbulent wrinkling caused by instabilities, rushing through the star. The flame propagation thus is faster than any hydrodynamic adjustment time scale, and an explosion is initiated, with little  expansion of white dwarf material. At densities around 10$^6$~g~cm$^{-3}$, this deflagration may turn into an explosion, and nuclear burning then competes with expansion of the material, resulting in some outer parts of the white dwarf not being burnt towards iron group nuclei, but only to intermediate-mass elements, or even left unburnt as carbon and oxygen mainly \citep{Mazzali:2007aa}. Typically, in a supernova of type Ia we expect that about 0.5 \Msol of nucleosynthetic \Ni are embedded in about 0.5--0.9~\Msol of other material \citep{Mazzali:2007aa,Stritzinger:2006aa,Scalzo:2014}. 

\Ni is unstable and decays first to \Co after $\tau\sim$~8 days, then  from \Co to $^{56}$Fe at $\tau\sim$~111 days. Initially, the supernova is expected to be still opaque to even gamma rays at MeV energies, converting this radioactive energy into emission at lower energy photons \citep{Clayton:1969aa,Hoeflich:1998ab}. As the supernova expands, more and more of the \Ni gamma-rays from radioactive decay thus should be able to leave the source region where the decay occurs, and be observable with gamma-ray telescopes \citep{Sim:2008aa,Isern:2008aa}. In particular during the second decay stage producing gamma-rays at 846.77 and 1238.29 keV, the supernova envelope is expected to become more and more transparent, so that radioactive decay and increasing transparency result in a maximum of gamma ray brightness at about 90--100 days after the explosion. As the supernova unfolds, the rise and fall of the gamma-ray line intensity provides unique information on the morphology of the inner ejecta as imprinted by the explosion and the restructuring of the supernova as radioactive energy is deposited in its thus-shaped interior \citep{Hoflich:2006a,Dessart:2014a}. This has been the objective for gamma-ray astronomy since these prospects have been proposed by \citet{Clayton:1969aa}.

SN2014J was discovered on January 22, 2014 \citep{Fossey:2014aa} in the nearby starburst galaxy M82 at \about~3.3 Mpc distance \citep{Foley:2014}as a type Ia explosion \citep{Cao:2014aa}. The explosion date appears to be 14 January, UT 14.75, with 0.2 to 0.3~d uncertainty \citep{Zheng:2014aa,Goobar:2015}. The supernova brightness maximum (blue band) was reached about 20 days after the explosion \citep{Goobar:2014aa}. Many studies from radio through infrared, optical and X-ray wavelengths were initiated, to study hints for its progenitor in pre-explosion data, hints for a companion star in emission details, and to follow the evolution of supernova light and spectra.

\section{INTEGRAL and SN2014J}
The INTEGRAL space gamma-ray observatory of ESA \citep{Winkler:2003} observed SN2014J from end January until end June, accumulating about 7~Msec of total exposure; some other brief observations interrupted this SN2014J campaign \citep{Kuulkers:2014}, and finally visibility constraints ended SN2014J observations. 
 
The ESA International Gamma-Ray Astrophysics Laboratory 'INTEGRAL' was launched into space in 2002, and carries two main telescopes, one called 'IBIS' \citep{Ubertini:2003} and optimised for imaging, and one called 'SPI'  \citep{Vedrenne:2003} and optimised as a gamma-ray spectrometer. Both INTEGRAL main telescopes use the \emph{coded mask} technique for imaging gamma-ray sources: a mask with occulting tungsten blocks and holes in the field of view of the gamma-ray camera casts a shadow from a celestial source onto the multi-element detector plane, which in the case of SPI consists of 19 Ge detectors, packed densely and with hexagonal shape. The SPI camera thus can resolve sources with a precision of 3~degrees, and its high-resolution Ge detectors obtain a spectrum at few keV resolution of celestial gamma-ray sources. The coded-mask shadowgrams, however, have to be recognised above a large instrumental background caused by cosmic ray interactions in the spacecraft and instrument materials. 

The data measured with SPI consist of energy-binned count spectra for the 15 of its 19 Ge detectors of the SPI telescope camera which were operational during our observations of SN2014J. 
The campaign for SN2014J of the INTEGRAL mission involved orbit numbers 1380 to 1428, mostly dedicated to SN2014J, with several short interruptions for technical reasons or monitoring of other sources, and one major gap between 23 April and 27 May. These observations were planned to cover mainly the rising part of the expected gamma-ray line emission, combined with a longer exposure at late times, when gamma rays should not be absorbed any more by supernova material. Then the total amount of \Co would be measured as it decays. Initially, we were in doubt as to where best invest the 'target of opportunity' time, considering that with even 2 Ms of observations on SN2011fe (distance about 6.5 Mpc) we had not achieved any hints for gamma-ray lines \citep{Isern:2013aa}. But  a few days after its discovery the supernova type and distance of about 3.5 Mpc were clear, and we decided to start observing on January 31, which was 16.3 days after the explosion \citep{Kuulkers:2014}. Recognising hints for the expected lines in quick-look data, we were able to convince the INTEGRAL user group and time allocation committee to continue monitoring of the supernova until visibility constraints terminated this opportunity on 26 June, 164.0 after the explosion. 

INTEGRAL observations are typically made as {3000-second} long pointings of the telescope towards a particular direction in the sky region of interest, moving the telescope axis by 2.1 degrees to shift the shadowgram of the source in the detector plane for the next 3000-second set, and so on, finally collecting exposure of the supernova in a regular pattern of telescope pointings of a 5 by 5 rectangle around the direction of SN2014J. The field of view of SPI is \about 30 degrees of opening angle, and always saw the source during those observations, at varying aspect angles. In total, after cleaning for data contaminated by solar flares or other irregularities, we collected 1816 telescope pointings, each with 15 detector spectra, hence 27240 spectra, each covering the 20-2000~keV energy band at 0.5~keV bin width. 

Our analysis method is based on a comparison of measured data to models in the complete set of count spectra as observed. For that, we must convert the expected shadowgrams for SN2014J in our set of pointings into the expected SN2014J counts using the SPI imaging response function, and we must develop a model for the large underlying instrumental background. 
We then fit the intensity scaling factor of the expected supernova contribution, plus a set of scaling factors per time for  the instrumental background, to the set of measured spectra. 

Our model for the instrumental background is derived from a detailed spectroscopic assessment of the long term background and detector behaviour, in which we make use of the various characteristic instrumental lines as they reflect isotopes and their de-excitation gamma-rays, modulated by degradations of the spectral resolution of each of our detectors and their recoveries through periodic annealing operations. Here we can account for the physical nature of instrumental background lines and of detector-specific spectral responses, and compared to earlier analyses we now combine data across a broader range of energy and time periods suitably to build a self-consistent description of instrumental background and its variations. Continuum and line contributions to the background are separately determined, detector responses and their degradations are determined from a combination of spectral lines and their longterm behaviour, respecting each detectors specific response, but combining signals for statistical precision as possible. We find that many aspects of the instrumental background can be understood and constrained from modeling a broader spectral range of typically 100 keV width. This overcomes limitations of modeling each energy bin separately from all others, as done before. 

We have analysed the first part of data in particular to search for early appearance of characteristic lines from \Ni decay, as this would provide information of how close \Ni may have been to the surface of the supernova. \Ni decay occurs at $\tau$=8.8~days, and thus the first few weeks of data are suitable to address this question. The results of this analysis have been published in \citet{Diehl:2014}. The full dataset then has been analysed to search for the appearance of the \Co decay lines as the supernova becomes more and more transparent. This decay at $\tau$=111~days occurs at the time scale at which the supernova reaches gamma-ray transparency, and the light curves and spectral shapes in the characteristic decay lines are the objective of this study, which have meanwhile also been published \citep{Diehl:2015}. In the following two Sections, we summarise and discuss both these studies, as they were presented at the AG conference in Bamberg in September 2014.

\section{\Ni near the surface and its early $\gamma$-rays}
\begin{figure}
  \centering
  \includegraphics[width=\linewidth]{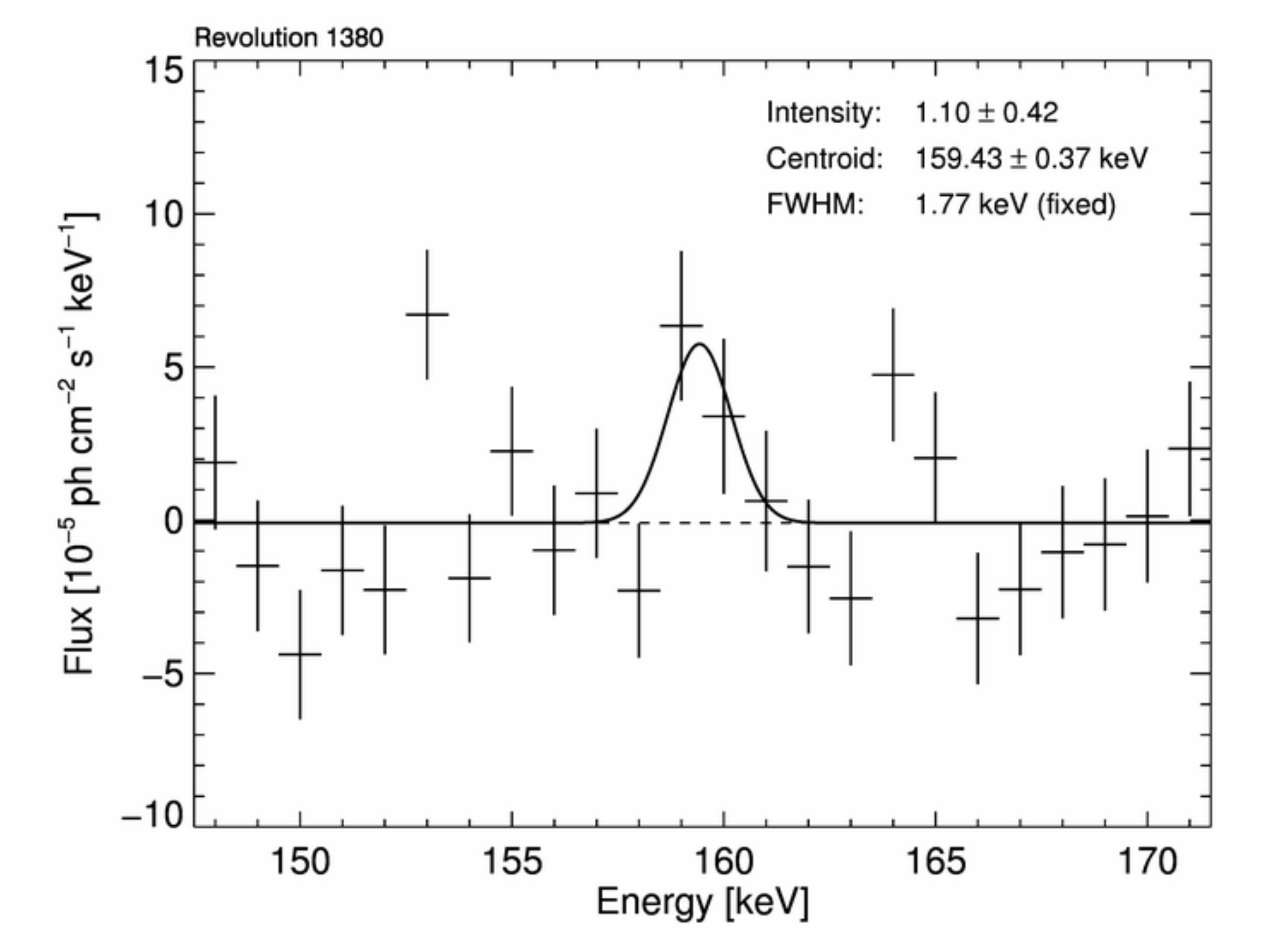}
   \caption{Gamma-ray spectrum measured with SPI/INTEGRAL from SN2014J. The observed three-day interval around day 17.5 after the explosion shows the two main lines from \Ni decay. In deriving these spectra, we adopt the known position of SN2014J, and use the instrumental response and background model. Error bars are shown as 1$\sigma$. The measured intensity corresponds to an initially-synthesised \Ni mass of 0.06 \Msol.}
  \label{fig:SPI-spectrum-158}
\end{figure}

\begin{figure}
  \centering
  \includegraphics[width=\linewidth]{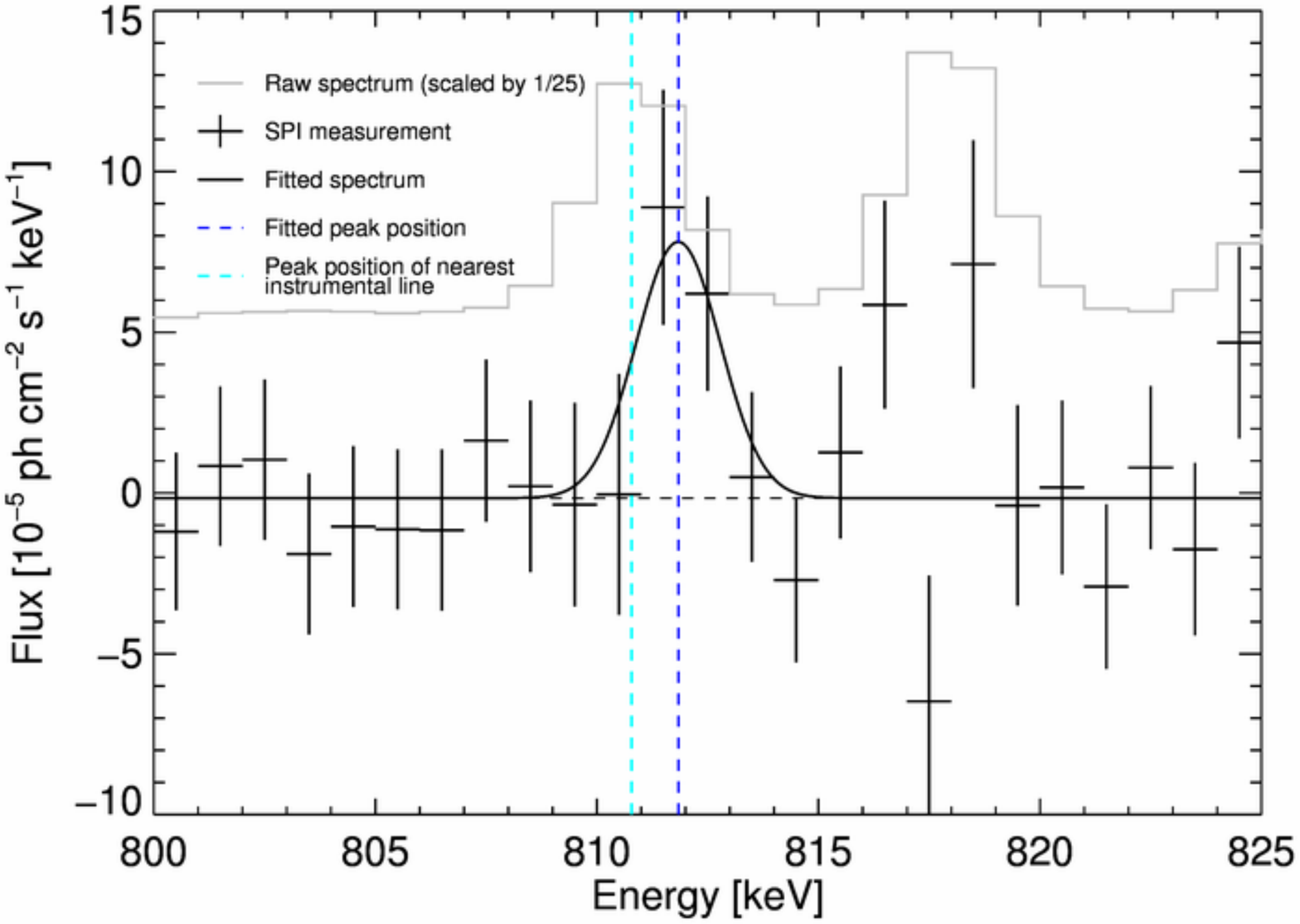}
   \caption{ Same as above, for the 812 keV line. We also illustrate our discrimination of sky and instrumental background, showing the SN2014J spectrum against a scaled raw data spectrum dominated by instrumental background lines. Evidently, the line from SN2014J appears offset from the centroid of a strong background line, but may be affected by it, as shown by the two high data points found at the position of another strong background line. Note, however, that the SN2014J line more-consistently follows the Gaussian response of SPI to a true gamma-ray line.}
  \label{fig:SPI-raw-fit-812}
\end{figure}

\begin{figure}
  \centering
  \includegraphics[width=\linewidth]{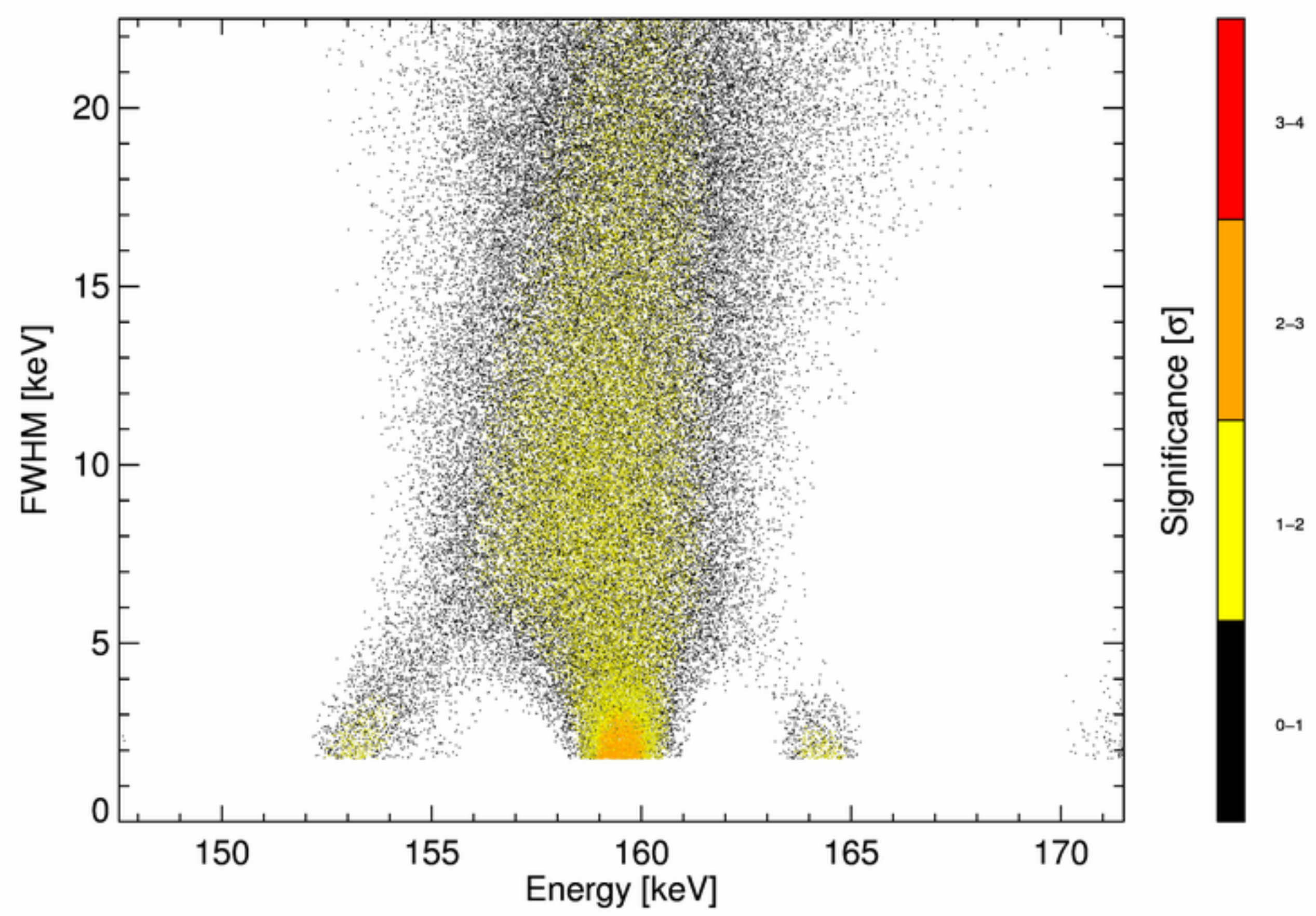}
   \caption{Illustration of line position and shape significances. Here we re-analyse the spectrum shown above with different parameters for centroid and width, plotting the line significance for each such trial. Evidently, the narrow and unshifted 158 keV line is the most probable result, but a lingle broad line as well as two other satellite lines may also be possible signals hidden in our data.}
  \label{fig:SPI-sign-158}
\end{figure}

\begin{figure}
  \centering
  \includegraphics[width=\linewidth]{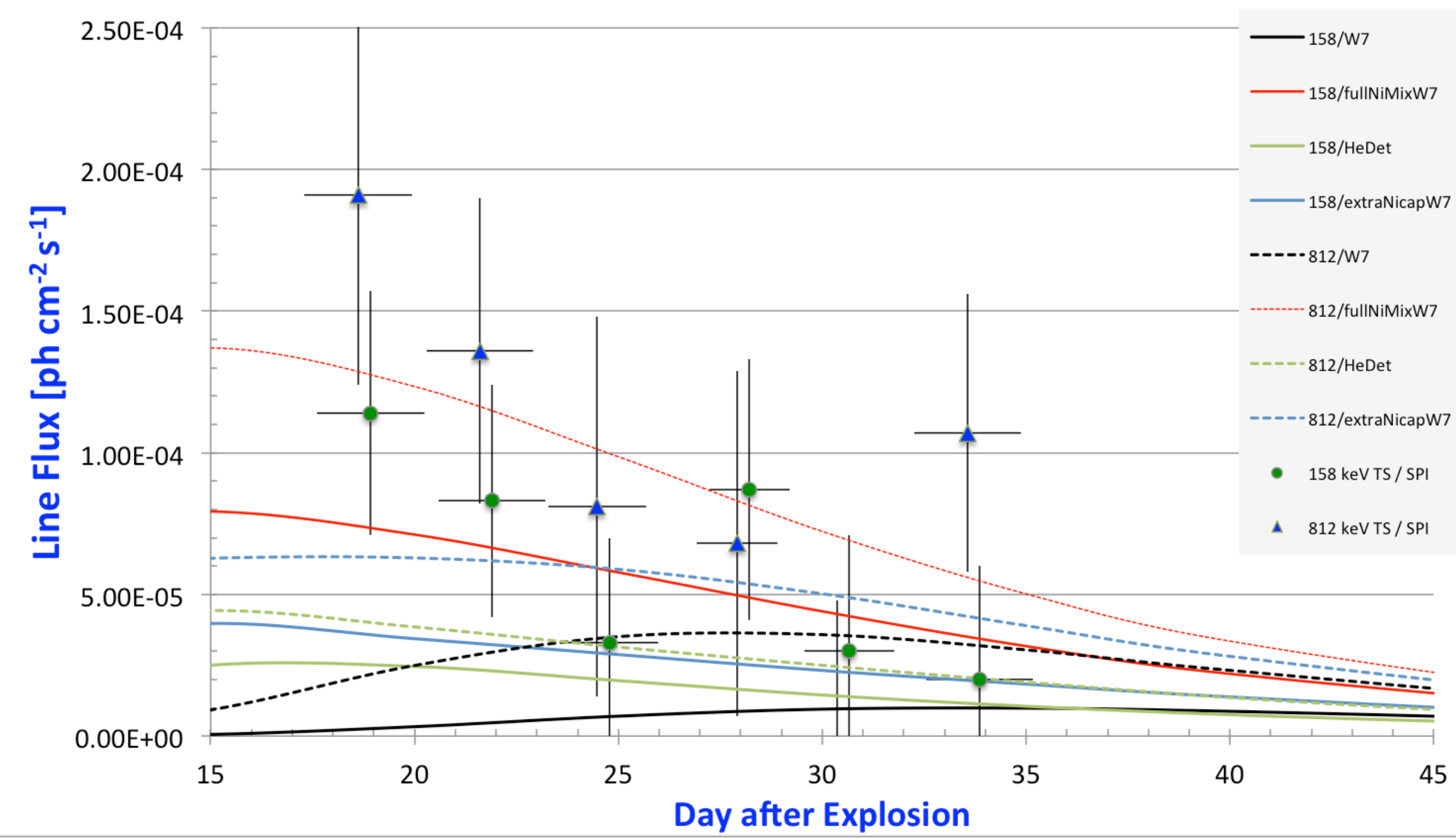}
   \caption{The fading of both lines from \Ni is consistent with the 8.8 day decay time, although statistical precision is poor, and observations should have better started earlier. We compare or measured intensity variation in the early period with several models from the set provided by \citet{The:2014}. Models with \Ni on or near the surface appear a better match than the canonical 'W7' model, although this remains speculative.}
  \label{fig:SPI-Nidecay}
\end{figure}

SN2014J observations with INTEGRAL started about two decay lifetimes of \Ni after the supernova explosion. It was therefore a great surprise and unexpected to see the two strongest lines from \Ni decay, as shown in Fig.\ref{fig:SPI-spectrum-158} to \ref{fig:SPI-Nidecay}. Both lines appear nearly at their laboratory energy values, line shifts are constrained to below 2100~km~s$^{-1}$. Also line broadening is modest and constrained to below 6000~km~s$^{-1}$, indicating that the \Ni near the surface does not expand with the higher velocities characteristic for outer supernova material, thus providing a hint for its possibly different origin.

This spectacular finding is puzzling in view of standard deflagration-detonation models. We therefore scrutinised our data and analysis methods, to investigate possible systematic uncertainties, given that the statistical significance of both lines is just at the 3$\sigma$ level. We show in Fig.~\ref{fig:SPI-raw-fit-812} that instrumental background lines present a major analysis challenge, but our modeling of instrumental background appears to properly account for them, within expectations. But instrumental background dominates the flux uncertainties in each spectral bin of our result, and both statistical and systematic limitations may allow for other interpretations. Testing this (see Fig.~\ref{fig:SPI-sign-158}), we cannot exclude that we might have been fooled by statistical excursions and the truth could also be a broader line shape, or multiple line components; we just report the most likely interpretation, exploiting with confidence the spectral capabilities of our instrument, and properly accounting for Poissonian statistical fluctuations as we fit our spectra. Finally, we point out that both lines at 158 and 812 keV are independent measurements, and are seen to fade in intensity consistent with what would be expected for \Ni decay ($\tau$=8.8~days) (see Fig.~\ref{fig:SPI-Nidecay}).

How can we make sense of this surprising result?
A single-degenerate Chandrasekhar-mass scenario appears unlikely to us: X-ray flux limits \citep{Nielsen:2014ac} and pre-explosion images \citep{Kelly:2014aa} exclude a super-soft progenitor. A sub-Chandrasekhar mass model with a He donor, or a merger of two white dwarfs, may seem better models for SN2014J, and could also explain our observation of \Ni in the outer layer of the supernova more readily. Moreover, this He donor progenitor channel is favored for this SN Ia from population-synthesis / supernova rate arguments \citep{Ruiter:2014}.  Also a classical double-detonation explosion scenario \citep{Fink:2010aa,Moll:2014}  is inconsistent with our observations: a \Ni shell engulfing the SN ejecta would be expected, resulting in broad, high-velocity gamma-ray emission lines, and moreover such an outer shell is expected to have an imprint on optical and infrared observables, which are probably not seen in SN2014J \citep{Goobar:2014,Telesco:2015}.

\begin{figure}
  \centering
  \includegraphics[width=\linewidth]{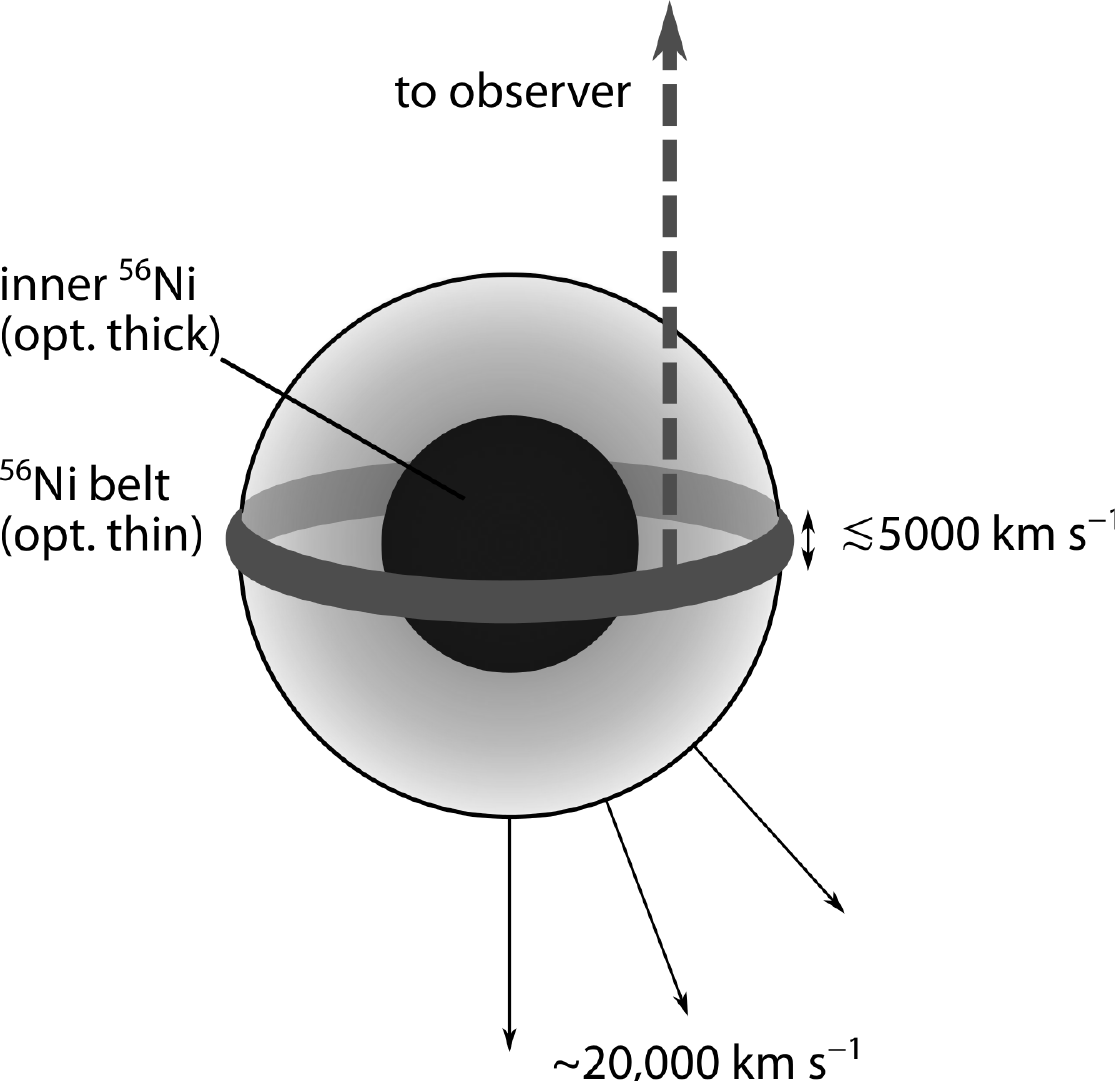}
   \caption{Sketch of a ejecta configuration compatible with our observations. Helium may have been accreted in a belt before the explosion, producing a \Ni-rich belt at the surface of the ejecta. The gamma-rays can escape early from the belt material, while the \Ni embedded more deeply and created by the main explosion (black) is still at high optical depths an invisible. A dashed arrow points to the observer, reflecting a face-on aspect within about 45$^{\circ}$ as required by a small if any shift of the observed line centroid. }
  \label{fig:Ni-belt}
\end{figure}

A modified version of such model may, however, guide us towards what might have happened (see Fig.~\ref{fig:Ni-belt}): If He would be accreted rapidly and form an equatorial accretion belt before it detonates, instead of accumulating in a shell, the kinematic constraints could be met, provided we observe the binary system essentially perpendicular to its orbital plane. Such an idea had been discussed frequently in the context of classical novae \citep{Kippenhahn:1978aa,Piro:2004ab,Law:1983}. This might be consistent with the observed gamma-ray signal, and compatible with optical observables. Our radiation transfer simulations in UV/optical/NIR  \citep[see][based on radiation transport, \citep{Kerzendorf:2014}]{Diehl:2014} shows that the Ni-belt would not produce easily distinguishable features but result in normal SN Ia appearance, not only for a pole-on observer but also for an equatorial observer. In view of this, our interpretation of an externally-triggered explosion may be plausible, though speculative. Further observables need to be checked against such a type of model.

\section{How $\gamma$-rays from \Co decay are revealed}
Clearly, in all variants of supernova type Ia models it is quite likely that major amounts of \Ni are produced, and also significant amounts of unburnt material are ejected, which occult gamma-rays from radioactive decay in the first months, as the supernova is spreading out. Unclear remains if \Ni is mixed throughout the supernova in the explosion, and hence some of it may already appear near the observable surface at early stages; this has been suggested by 3-dimensional hydrodynamical simulations \citep{Seitenzahl:2013aa}. Also unclear remains if \Ni production in the central parts is reduced due to electron capture dominance at high densities, as suggested by flattened line profiles of Co lines that can be seen in the mid infrared \citep{Gerardy:2007a}. All occultation of gamma rays should decrease with time and be rather unimportant after about three months. 

\begin{figure}
  \centering
  \includegraphics[width=\linewidth]{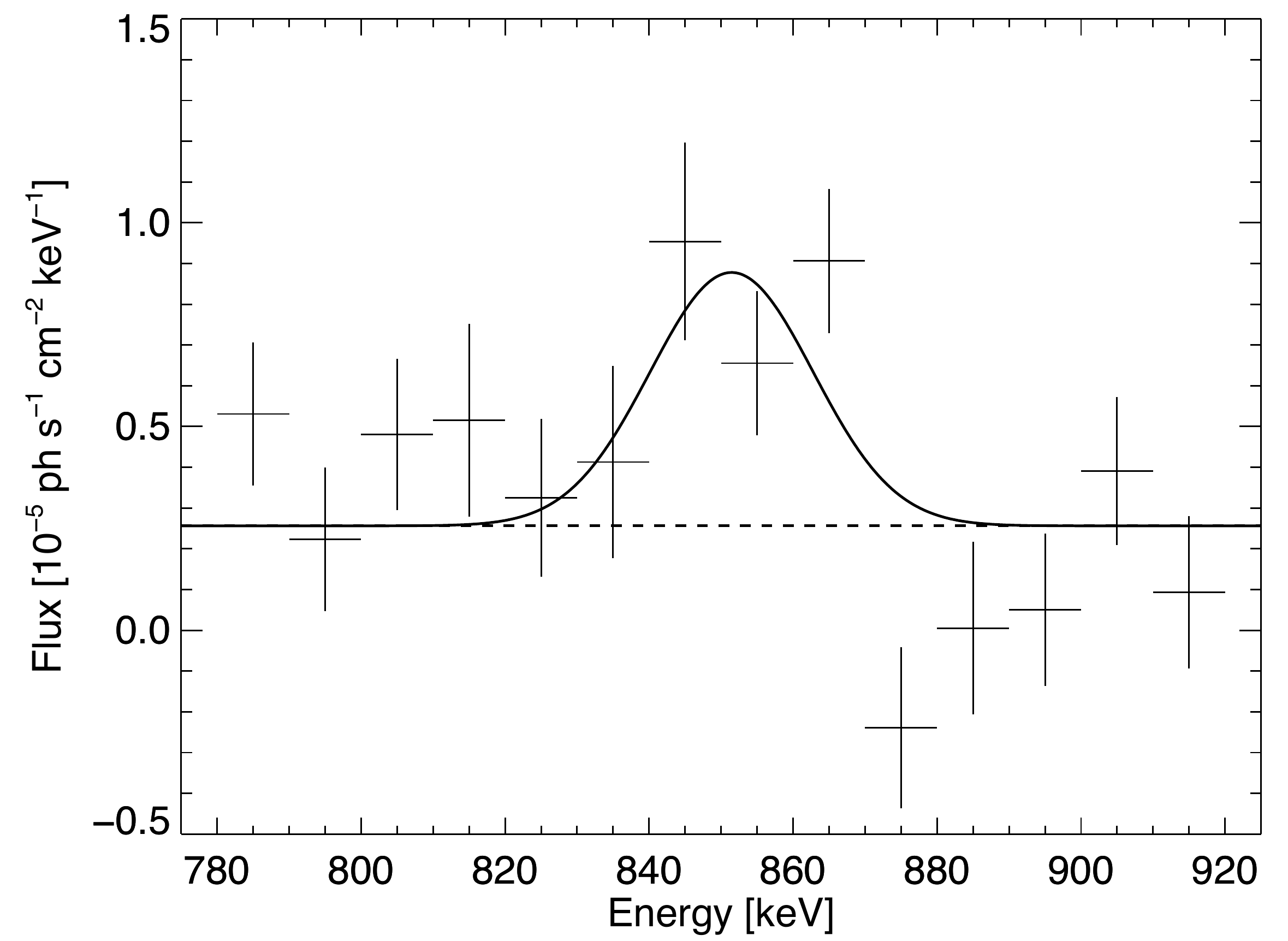}
  \includegraphics[width=\linewidth]{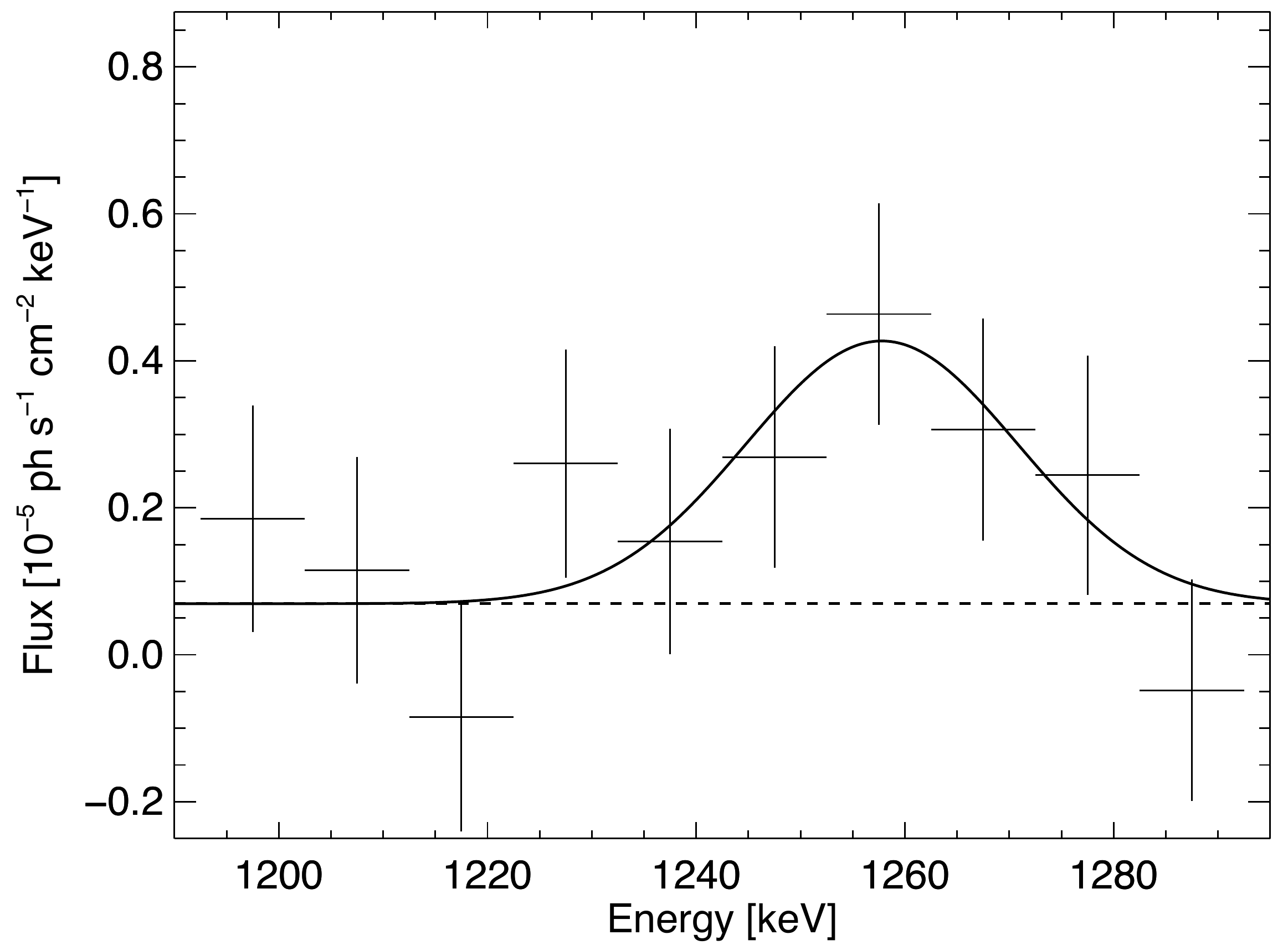}
   \caption{SN2014J spectrum near the 847 keV line ({\it above}) and near the 1238 keV line ({\it below}) as expected from \Co decay. These spectra are determined in energy bins  of width 10 keV over the entire observing period; the source intensity is fitted at four independent epochs. For illustration, fitted Gaussians indicate the detection of broadened lines near the \Co gamma-ray line energies.}
  \label{fig:SPI-spectrum-847-1238}
\end{figure}

Other INTEGRAL data analysis results had reported \Co decay with its strongest lines at 847 and 1238 keV to appear as broad lines, quite in concordance with standard Chandrasekhar mass models \citep{Churazov:2014b}. Our own analysis of time integrated data confirms this general picture (see Fig.~\ref{fig:SPI-spectrum-847-1238}). 

\begin{figure}
  \centering
  \includegraphics[width=\linewidth]{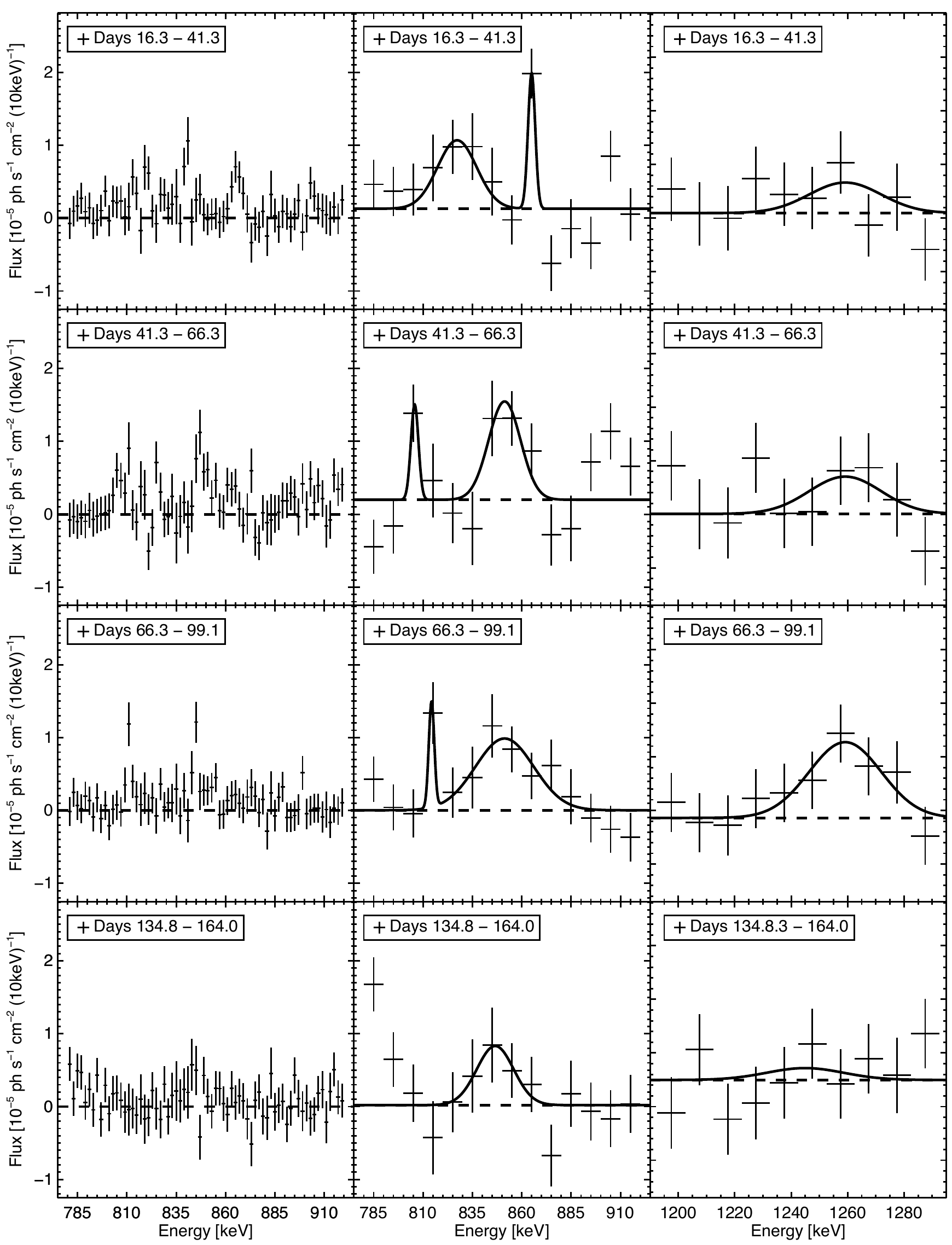}
   \caption{SN2014J signal intensity variations for the 847 keV line ({\it center}) and  the 1238 keV line ({\it right}) as seen in the four epochs of our observations, and analyzed with 10 keV energy bins. The 1238 keV fluxes have been scaled by the \Co decay branching ratio of 0.68 for equal-intensity appearance. Clear and significant emission is seen in the lower energy band ({\it left and center}) through a dominating broad line attributed to 847~keV emission, the emission in the high-energy band in the 1238~keV line appears consistent and weaker, as expected from the branching ratio of 0.68  ({\it right}). Fitted line details are discussed in the text.
   For the 847 keV line, in addition a high-spectral resolution analysis is shown at 2 keV energy bin width ({\it left}), confirming the irregular, non-broad-Gaussian features in more detail.  }
  \label{fig:SPI-spectra-set-847-1238}
\end{figure}

But as we exploit the full spectral resolution of our instrument, we find that these broad lines seen at late epochs are not quite the same (though just occulted and at lower brightness) towards earlier epochs, see Fig.~\ref{fig:SPI-spectra-set-847-1238}. They may, in fact, be composed of narrower emission lines which vary in intensity as the supernova unfolds. This could reflect emission from a few \Co-rich plumes, embedded in the supernova and thus progressively revealed through structured non-radioactive ejecta with a more complex morphology.

\begin{figure}
  \centering
  \includegraphics[width=\linewidth]{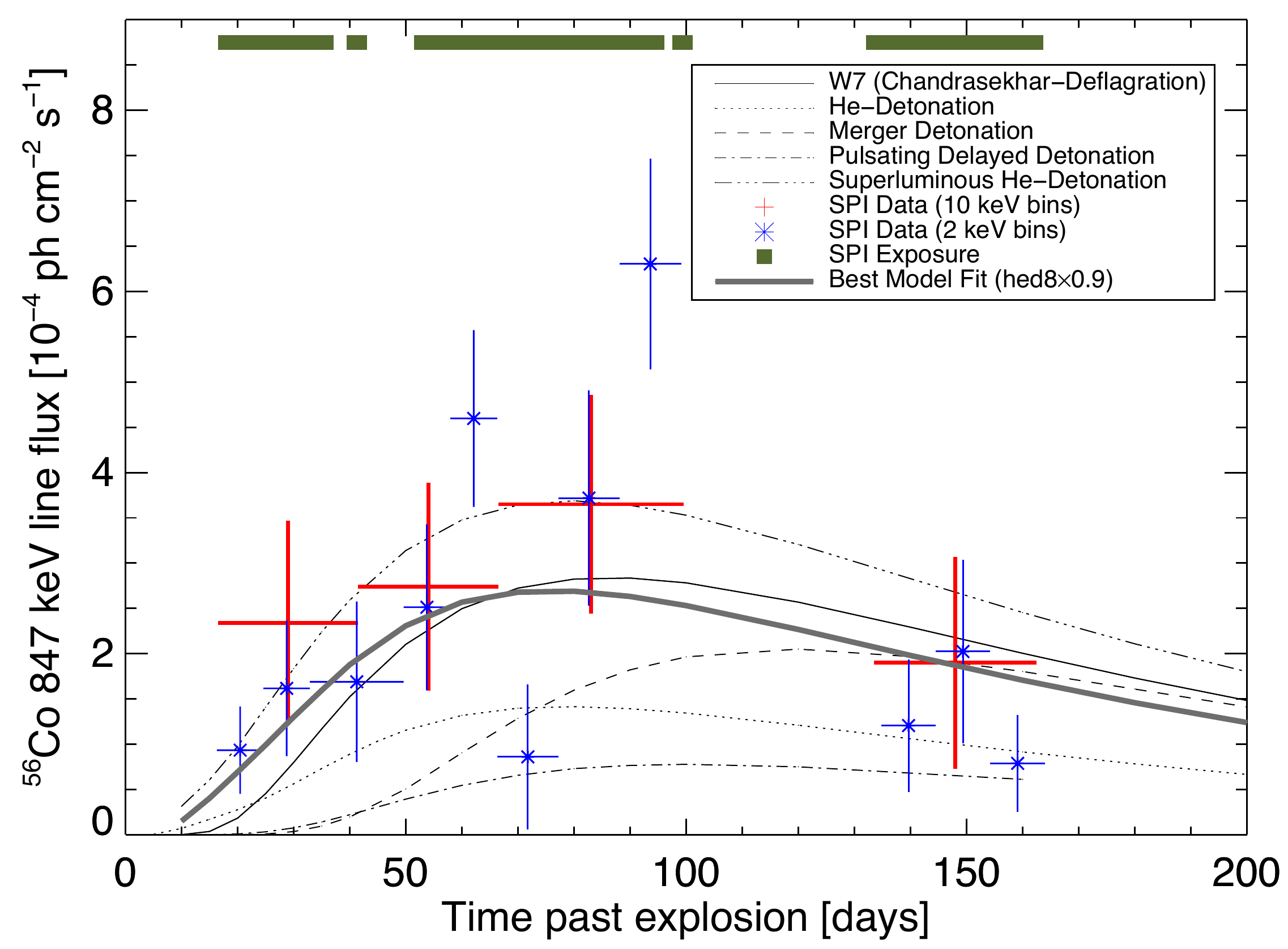}
   \caption{SN2014J signal intensity variations for the 847 keV line in two different time resolutions. The 4-epoch results are consistent with 11-epoch analysis, both showing an initial rise and late decline of \Co decay line intensity, with a maximum at 60--100 days after explosion. Shown are also several candidate source models from  \citet{The:2014}, fitted in intensity and thus determining \Ni masses for each such model. The best-fitting model is shown as a continuous thick line. The \Ni mass has been derived from such model fits as 0.49$\pm$0.09~\Msol \citep[see][for details]{Diehl:2015}}
  \label{fig:SPI-lightcurve-847-fine}
\end{figure}

It is difficult to cut our observations into as many spectral and timing bins as we would like, to disentangle the signature from how \Co may be embeded in the supernova. Attempting to maximise time resolution without imposing a bias from a particular supernova model or simulation, we generated a light curve from our data alone, in the more-significant signal of the 847 keV line (see Fig.~\ref{fig:SPI-lightcurve-847-fine}). In this Figure, we show for comparison also the four epochs where Figure \ref{fig:SPI-spectra-set-847-1238} shows the detailed spectra and their intensity variations seen more significantly. Clearly, this borders on what can be extracted given the uncertainty of our measurements; but our aim is to learn from the specific messages of gamma-ray data. Alternatively, if 3-dimensional hydrodynamic models could be combined with current sophisticated radiation transport models, we could compare such more flexible 3D model-predicted gamma-ray light curves to our observations. This is, however, beyond the scope of current simulations and analysis. Our analysis shows that significant variability characterises how gamma rays from \Co decay appear, as time goes on and the supernova ejecta become gamma-ray transparent, clearly beyond the smooth light curves currently available from supernova model predictions. More details can be found in \citet{Diehl:2015}.

\section{Summary and conclusions}
Supernova SN2014J is the first supernova of type Ia which is close enough for significant measurements of characteristic gamma rays from the \Ni decay chain. INTEGRAL has followed the gamma-ray emission for five months, and thus obtained the first clear detection of such characteristic gamma ray emission. This provides the first direct confirmation that radioactive decay from \Ni through \Co is the energy source of supernova light, and that about 0.5~\Msol of \Ni have been seen in gamma-rays from SN2014J. Gamma-ray data directly from \Ni radioactive decay have finally confirmed a key aspect of our understanding of type Ia supernovae.

Upon a closer look, there are challenges and surprises, presented by those same gamma ray observations: Early time \Ni gamma-ray emission has been found, and is surprising in appearance. Its brightness suggests a major fraction of near 10\% of the total \Ni be visible 17 days after the explosion. It is unclear if this is \Ni contributed by a surface event, or if this happened to be a \Ni rich plume rising early from the inner supernova region to the surface. The later gamma-ray spectra and their \Co decay lines then show  structure that may raise significant doubts about a homogeneous and smooth distribution of \Ni throughout the supernova. It remains a challenge how these \Co decay data can be reconciled with data from the receding photosphere at lower wavelength regimes, which altogether build a tomographic view of the angle-averaged  morphology of the supernova. SN2014J provides a challenge to our understanding of type Ia supernovae, on second glance. We will learn if that remains a specific and puzzling event, or if underlying processes linked to \Ni radioactive decay may not be apparent in data of stable ejecta.

\acknowledgements
This highlight paper is based on work of our MPE team analyzing INTEGRAL/SPI data, with Thomas Siegert, Martin Krause, Jochen Greiner, Wei Wang, and Xiaoling Zhang, and on supernova physics discussions and simulations with Wolfgang Hillebrandt, Markus Kromer, Keiichi Maeda, Friedrich R\"opke, Stuart Sim, and Stefan Taubenberger. We appreciate discussions on the subjects of this paper in particular with Adam Burrows, Eugene Churazov, Sergei Grebenev, Peter H\"oflich, Jordi Isern, Paolo Mazzali, R\"udiger Pakmor, Jean-Pierre Roques, Ivo Seitenzahl, and many others. 
     
       This research was supported by the German DFG cluster of excellence 'Origin and Structure of the Universe', and from DFG Transregio Project No. 33 'Dark Universe'. F.K.R. was supported by the DFG (Emmy Noether Programm RO3676/1-1) and the ARCHES prize of the German Ministry for Education and Research (BMBF). The work by K.M. is partly supported by JSPS Grant-in-Aid for Scientific Research (23740141, 26800100) and by WPI initiative, MEXT, Japan. W.W. is supported by the Chinese Academy of Sciences and the Max Planck Gesellschaft, and by the German DFG cluster of excellence 'Origin and Structure of the Universe'. We are grateful to E. Kuulkers for handling the observations for the INTEGRAL SN2014J campaign.
  The INTEGRAL/SPI project
  has been completed under the responsibility and leadership of CNES;
  we are grateful to ASI, CEA, CNES, DLR, ESA, INTA, NASA and OSTC for
  support of this ESA space science mission.
  INTEGRAL's data archive (http://www.isdc.unige.ch/integral/archive\#DataRelease) is at the ISDC in Versoix, CH, and includes the SN2014J data used in this paper.

\bibliographystyle{aa}
\bibliography{rod-refs_SNIa,rod-refs13}

\end{document}